\documentclass[letterpaper,twocolumn,10pt]{article}
\usepackage{usenix2019_v3}
\usepackage{enumitem}
\usepackage[utf8]{inputenc}
\usepackage[english]{babel}
\usepackage{booktabs}
\usepackage{arydshln}  % Add this to your preamble

\usepackage{tcolorbox}
\tcbuselibrary{skins} % This loads the "enhanced" skin
\tcbuselibrary{listings}

\usepackage{tikz}
\usepackage{amsmath}

\usepackage{filecontents}

\newcommand{\attack}{Spill The Beans}

\usepackage{fancyhdr}
\fancypagestyle{firstpage}{
  \fancyhf{} % clear all header and footer fields
  
  \fancyfoot[C]{%
    \parbox{\textwidth}{%
      \centering % This centers the content within the parbox
      Approved for Public Release; Distribution Unlimited. Public Release Case Number 25-0107 \\
      ©2024 The MITRE Corporation. ALL RIGHTS RESERVED. - \thepage%
    }%
  }
}    

\begin{document}
%-------------------------------------------------------------------------------

%don't want date printed
\date{}

\title{\Large \bf \attack : Exploiting CPU Cache Side-Channels to Leak Tokens from \\
Large Language Models}

%for single author (just remove % characters)
\author{
{\rm Andrew Adiletta}\\
MITRE
\and
{\rm Berk Sunar}\\
Worcester Polytechnic Institute
% copy the following lines to add more authors
% \and
% {\rm Name}\\
%Name Institution
} % end author

\maketitle
\thispagestyle{firstpage}

%-------------------------------------------------------------------------------
\begin{abstract}
%-------------------------------------------------------------------------------

Side-channel attacks on shared hardware resources increasingly threaten confidentiality, especially with the rise of Large Language Models (LLMs). In this work, we introduce \attack, a novel application of cache side-channels to leak tokens generated by an LLM. By co-locating an attack process on the same hardware as the victim model, we flush and reload embedding vectors from the embedding layer, where each token corresponds to a unique embedding vector. When accessed during token generation, it results in a cache hit detectable by our attack on shared lower-level caches.

A significant challenge is the massive size of LLMs, which, by nature of their compute intensive operation, quickly evicts embedding vectors from the cache. We address this by balancing the number of tokens monitored against the amount of information leaked. Monitoring more tokens increases potential vocabulary leakage but raises the chance of missing cache hits due to eviction; monitoring fewer tokens improves detection reliability but limits vocabulary coverage.

Through extensive experimentation, we demonstrate the feasibility of leaking tokens from LLMs via cache side-channels. Our findings reveal a new vulnerability in LLM deployments, highlighting that even sophisticated models are susceptible to traditional side-channel attacks. We discuss the implications for privacy and security in LLM-serving infrastructures and suggest considerations for mitigating such threats. For proof of concept we consider two concrete attack scenarios: Our experiments show that an attacker can recover as much as 80\%-90\% of a high entropy API key with single shot monitoring. As for English text we can reach a 40\% recovery rate with a single shot. We should note that the rate highly depends on the monitored token set and these rates can be improved by targeting more specialized output domains.
\end{abstract}

%-------------------------------------------------------------------------------
\section{Introduction}
%-------------------------------------------------------------------------------

The widespread deployment of Large Language Models (LLMs) has revolutionized natural language processing tasks, enabling applications in various domains such as finance, healthcare, and customer service. However, the increasing reliance on these models raises significant concerns about privacy and security, especially in multi-user environments where shared hardware resources can be exploited. 

%\subsection{Prior Work on Side-Channel Attacks in LLMs}

\paragraph{Attacks on Key-Value (KV) Caches.}
Recent studies have identified vulnerabilities in LLM inference systems arising from the use of caching mechanisms such as Key-Value (KV) caches~\cite{song2024early, zheng2024inputsnatch}. KV caches store intermediate computations, specifically the attention key and value vectors, to accelerate inference by reusing them for similar input sequences. However, subtle timing differences emerge when these caches either hit or miss, allowing attackers to infer sensitive data based on how quickly the LLM responds.

Song \textit{et al.}~\cite{song2024early} showed that by closely measuring the time taken to generate outputs, it is possible to detect cache hits in KV caches, thereby reconstructing private prompt tokens or system instructions. Similarly, Zheng \textit{et al.}~\cite{zheng2024inputsnatch} described a timing-based side-channel approach to compromise LLM inference by manipulating input candidates and carefully analyzing response latencies. The ability to leak entire conversations through KV caches has spurred research into software-level defenses. A recently published work~\cite{li2024coreguard} proposes \textit{KV-Shield}, which permutes weight matrices at initialization so that leaked KV pairs are effectively obfuscated. Their approach mitigates KV-based leakage by confining essential permutation operations within a trusted execution environment (TEE), thereby denying direct GPU access to unencrypted KV data.

\paragraph{Output Token Count and Network-Based Leakage.}
Other recent research spotlights the risks posed by timing and length-based side-channels that exploit how many tokens an LLM generates at inference time. In \cite{zhang2024time}, the authors show that measuring response latencies can reveal an LLM’s output token count, leaking sensitive attributes like the target language or output class with high accuracy. Their proposed attacks take advantage of token-count biases in multilingual translation tasks and text classification scenarios, and extend to remote, network-based timing measurements. Similarly, \cite{weiss2024your} demonstrates a token-length side-channel in real-time LLM responses, revealing entire sentences over encrypted browser or API connections. By leveraging a large language model to reconstruct plaintext based on observed token-length patterns, these attacks effectively perform remote keylogging of a victim’s AI assistant interactions. Although both efforts share our high-level motivation—uncovering hidden channels for token inference—our approach diverges by targeting the \textit{hardware-level} cache for LLM embeddings, whereas they rely on timing or length exposure within the application or network stack. Thus, existing mitigations aimed at attenuating token-count or response-length signals may not directly thwart low-level attacks on shared CPU caches.

\paragraph{Hardware Cache vs.\ KV Cache Leakage.}
Despite these efforts to secure software-level KV caches, no prior work (to our knowledge) focuses on microarchitectural caches in the LLM inference pipeline. Existing KV cache defenses do not address an attack that occurs below the software layer, specifically targeting the CPU’s cache hierarchy where embedding matrices and other model parameters reside. In contrast to KV caches—which are generally managed in user-space and can be directly patched or permuted—hardware caches are governed by the CPU microarchitecture, making them substantially harder to protect through conventional software approaches.

LLMs process input tokens by mapping them to embedding vectors in the embedding layer~\cite{bengio2000neural, mikolov2013distributed}. This mapping transforms high-dimensional, sparse data into a dense, lower-dimensional space where semantic relationships between tokens are captured. The embedding layer serves as a lookup table where each unique token corresponds to a specific embedding vector. During inference, when a token is processed, its corresponding embedding vector is retrieved from the embedding layer to generate contextually appropriate responses. In multi-user systems or cloud environments, these embedding matrices may reside in shared caches, creating a potential avenue for side-channel attacks. Since the embedding layer accesses specific memory locations corresponding to the tokens being processed, an attacker co-located on the same hardware can potentially monitor cache access patterns to infer the tokens used by the victim LLM.

The universality of cache hierarchies and common memory-optimization features (like page deduplication) create fertile ground for side-channel attacks. An attacker executing any of these techniques can potentially extract high-resolution information about a victim’s data usage, from cryptographic operations to natural-language tokens in an LLM-serving environment. Mitigations typically center on preventing fine-grained attacker observations —-either by disallowing shared pages, masking memory access patterns, or implementing hardware cache partitioning—- yet these remedies often impose unacceptable performance or deployment overhead, making cache side-channels a persistent threat. 

This work thus is the first to demonstrate an attack on LLMs that leverages a hardware-level cache side-channel (e.g., via Flush+Reload \cite{yarom2014flush}). While KV-Shield, FHE, or TEE-based protocols might mitigate KV leakage, they do not naturally extend to embedding-layer accesses or other parameters stored in physical caches. Our approach illustrates that even if software-layer caches are secured, sensitive tokens can still be exfiltrated by monitoring lower-level memory resources shared by attacker and victim processes. This leaves LLM deployments susceptible to hardware-level leaks, necessitating a broader defense strategy that addresses both KV-level and hardware-level side-channels.

\subsection*{Our Contribution}

In this paper, we introduce \textit{\attack}, a novel application of cache side-channels to attack to leak tokens produced by an LLM. Specifically, we:

\begin{itemize}[noitemsep,topsep=0pt,leftmargin=*]
\item Introduce a new attack vector targeting LLM outputs on co-located servers via cache accesses exploiting data coherency in unified CPU/GPU memory;
\item Unlike earlier side-channel attacks, our approach recovers tokens precisely even with a single measurement; repeated runs with the same prompt permit full recovery of the LLM output.
\item Depending on the topic of the LLM output, we can recover as much as 40\% of the tokens of plain English with a single shot. As for high entropy API keys, an attacker can recover as much as 80\%-90\% of the key with single shot monitoring since higher coverage rates are achievable with the restricted token set in API keys. 
\item Explain how the LLM embedding layer's access patterns can be exploited with CPU cache monitoring to infer the tokens being processed by the model;
\item Address the challenge of cache eviction in large models by experimenting with trade-offs with overhead and the number of tokens monitored and the resulting amount of information leaked;
\item Validate the effectiveness of \textit{\attack} through extensive experimentation, highlighting the feasibility of leaking tokens from LLMs via cache side-channels, and providing an end-to-end example where we leak a users' sensitive API key;
\item Discuss the overall security implications of shared hardware for LLM inferencing and potential mitigations.

\end{itemize}

%%%%%%%%%%%%%%%%%%%%%%%
\section{Background}
%%%%%%%%%%%%%%%%%%%%%%%
Modern computing systems employ hierarchical memory architectures to bridge the performance gap between fast processors and slower main memory. Understanding this memory hierarchy is important for analyzing cache attacks which exploit these architectural features to leak sensitive information. Additionally, the internal mechanisms of Large Language Models (LLMs), particularly their use of embedding layers, present unique vulnerabilities that can be targeted by such attacks.

\subsection{Memory Hierarchy and Cache Side-Channel Attacks}

\paragraph{Memory Hierarchy in Modern Processors.}
Contemporary CPUs are designed with a multi-level cache hierarchy to improve computational efficiency. This hierarchy typically includes small, fast Level 1 (L1) caches closest to the CPU cores, larger Level 2 (L2) caches, and even larger shared Level 3 (L3) caches~\cite{arch2023Hennessy}. The cache hierarchy serves to reduce the latency of memory accesses by storing frequently accessed data closer to the processor.

In multi-core processors, lower-level caches such as L3 are often shared among cores, enabling faster inter-process communication but also introducing potential security risks~\cite{yarom2014flush}. The shared nature of these caches allows processes running on different cores to influence each other's cache states, forming the basis for cache side-channel attacks across cores.

\paragraph{Timing Side-Channel Attacks.}
Timing side-channels exploit the fact that the time required to process certain operations often leaks sensitive information about the internal state of a system. By carefully measuring the execution latency of a target operation, an adversary can infer whether particular code paths were taken, which data structures were accessed, or which cryptographic keys were used. Early work on timing side-channels focused primarily on cryptographic routines~\cite{kocher1996timing, brumley2005remote}, where slight variations in exponentiation or multiplication time revealed secret keys to a remote attacker. Subsequent research broadened the scope to operating systems, hypervisors, and various microarchitectural resources such as caches, branch predictors, and DRAM row buffers~\cite{tromer2010efficient, osvik2006cache}.

In the broader landscape of side-channel attacks, timing analysis is especially potent due to its low cost and broad applicability. Unlike fault attacks or electromagnetic (EM) analysis, timing side-channels typically require no specialized hardware or physical proximity; measurements are often performed purely in software, either locally or remotely. When combined with other microarchitectural side-channels—such as Flush+Reload or Prime+Probe on shared CPU caches—timing measurements enable high-resolution observations of what should be private operations in a victim process~\cite{yarom2014flush, osvik2006cache}. These attacks can leak cryptographic keys, user keystrokes, web browsing histories, and, as our results show, sensitive language-model tokens. 

In multi-tenant environments such as public clouds, timing side-channels become even more concerning. Here, an attacker can co-locate with a victim by simply renting or deploying a virtual machine on the same physical hardware. Despite the isolation guarantees at the virtualization layer, shared hardware resources remain susceptible to precise timing probes that reveal inter-VM interactions~\cite{zhang2012cross}. Modern defenses often revolve around strategies like time partitioning, memory obfuscation, or hardware partitioning, but each faces practical deployment and performance trade-offs. Consequently, understanding and mitigating timing side-channel risks is critical across a wide array of platforms, from high-performance data centers to consumer-grade hardware.

\paragraph{Cache Side-Channel Attacks.}
Cache side-channel attacks exploit the timing differences between memory accesses that "hit" in the cache versus those that "miss" and require fetching data from main memory~\cite{tromer2010efficient}. By carefully measuring access times, an attacker can infer sensitive information about the victim’s memory access patterns. Techniques like Flush+Reload~\cite{yarom2014flush}, Prime+Probe~\cite{kayaalp2016high}, and Prime+Scope~\cite{purnal2021prime} have been used to extract cryptographic keys, keystroke information, and other confidential data with high accuracy.

Flush+Reload is particularly powerful because it offers fine-grained insight into which exact memory lines have been accessed by a victim process. It leverages shared pages between attacker and victim, often introduced via memory deduplication (e.g., Kernel Same-Page Merging in Linux) or shared libraries. To conduct the attack, the attacker first flushes a specific cache line throughout the cache hierarchy using instructions like \texttt{clflush} and subsequently times how long a reload takes. If the reload is fast, the attacker infers that the victim must have accessed the line in the interim, thereby re-caching it. This mechanism is enabled by the inclusive nature of many modern CPU cache architectures: when data is flushed from the last-level cache (LLC), it is also evicted from higher-level caches, ensuring uniform eviction across the hierarchy~\cite{yarom2014flush}. Techniques like Flush+Reload~\cite{yarom2014flush} have been extensively studied in the context of cryptographic implementations~\cite{gullasch2011cache}, enabling attackers to recover secret keys and other confidential data.

Prime+Probe and Prime+Scope, in contrast, do not require shared pages. Prime+Probe primes (i.e., fills) a specific cache set with attacker-controlled data and later probes those same cache lines to detect whether any were evicted by the victim’s accesses~\cite{kayaalp2016high}. Prime+Scope adds further refinement by zooming in on smaller cache slices or ways, offering a more efficient method to infer cache usage with reduced noise~\cite{purnal2021prime}. Although these techniques can provide near real-time visibility into a victim’s cache utilization, they often demand more detailed knowledge of cache indexing and associativity. Nonetheless, each approach capitalizes on the principle that cache activity—measurable by timing—betrays the precise memory addresses or cache sets involved in a victim’s computation.

\paragraph{Cache Attacks in Cloud Environments.}
Prior research has also explored cache side-channel attacks in cloud and multi-tenant environments~\cite{zhang2012cross}. These studies show that attackers can exploit shared caches to extract sensitive information from co-located virtual machines or processes. Such attacks pose a significant risk in cloud-based LLM deployments, where shared physical hardware among multiple users can be exploited to infer sensitive information, such as private inputs, outputs, or configuration details, from co-located processes.

\subsection{Page Deduplication and Memory Sharing}

\paragraph{Kernel Same-Page Merging (KSM).}
Page deduplication, commonly referred to as Kernel Same-Page Merging (KSM) in Linux systems, is a memory optimization technique that identifies identical memory pages across different processes or virtual machines (VMs) and merges them into a single physical page. This process reduces the overall memory footprint by eliminating redundant data, which is particularly beneficial in environments running multiple instances of similar applications or operating systems.

KSM operates by scanning the main memory for pages with identical content. Once such pages are found, they are consolidated into a single page, and all references to the original pages are updated to point to this shared page. The shared page is marked as copy-on-write (COW), ensuring that if any process attempts to modify the page, a private copy is created for that process to maintain data integrity.

\paragraph{Benefits in Virtualized Environments.}
In virtualization scenarios, where numerous VMs may run the same guest operating systems or applications, a significant amount of memory duplication occurs~\cite{waldspurger2002memory}. By employing page deduplication, hypervisors can greatly reduce memory usage, allowing for higher VM densities on a single physical host. KSM can enable running dozens of virtual instances with limited physical memory by sharing common pages among them.

\paragraph{Security Implications.}
While page deduplication offers memory efficiency benefits, it introduces security vulnerabilities that can be exploited by side-channel attacks~\cite{gruss2015practical}. The shared nature of deduplicated pages enables attackers to perform high-resolution cache side-channel attacks like Flush+Reload~\cite{yarom2014flush}. Since the attacker and victim processes share physical memory pages, the attacker can monitor cache access patterns to infer sensitive information from the victim.

Moreover, page deduplication can be abused to circumvent Address Space Layout Randomization (ASLR), a security mechanism that randomizes memory addresses to prevent exploitation~\cite{bosman2016dedup}. By detecting whether a page has been merged, attackers can gain insights into the memory layout of the victim process.

%%%%%%%%%%%%%%%%%%%%%%
\section{Related Work}
%%%%%%%%%%%%%%%%%%%%%%
\paragraph{Limitations of Prior Work.}
Previous attacks on LLMs have predominantly focused on software-level vulnerabilities, such as exploiting API behaviors, prompt manipulation, or timing discrepancies in application-layer caching mechanisms~\cite{song2024early}. These approaches often rely on specific software configurations or assumptions and typically extract only partial or semantic information rather than exact token sequences, and additionally are many-shot. In contrast, our work is the first to exploit hardware-level microarchitectural vulnerabilities to leak tokens from LLMs, one-shot. By targeting the embedding layer directly using the Flush+Reload technique, we are able to recover exact tokens with high precision. This hardware-based attack operates at a lower level than prior methods, bypassing software defenses and highlighting a previously unexplored vector for information leakage in LLMs.

%%%%%%%%%%%%%%%%%%%%%%%%%%%%%%%%%%%%%%%%%%%%
\section{\attack}
%%%%%%%%%%%%%%%%%%%%%%%%%%%%%%%%%%%%%%%%%%%%
%%%%%%%%%%%%%%%%%%%%%%%%%%%%%%%

\begin{comment}
\paragraph{Attack Overview} Figure~\ref{fig:attack_overview} depicts the \attack\ attack. We envision a malicious party runs a process monitoring cache accesses of a victim. The victim is running software that gives it access to an LLM. Typically the victim (and attacker) run on a CPU as standalone tasks possibly in isolated execution environments, while the inference is run on an attached GPU device. The attacker monitors accesses, e.g. using Flush+Reload, to the embedding vectors residing in the cache by the victim during the token embedding stage during an LLM inference. This allows the adversary to reconstruct the output of the victim's LLM inference token by token as they are decoded through the embedding vectors.
\end{comment}

\begin{figure}[h!]
    \centering
    \includegraphics[width=\columnwidth]{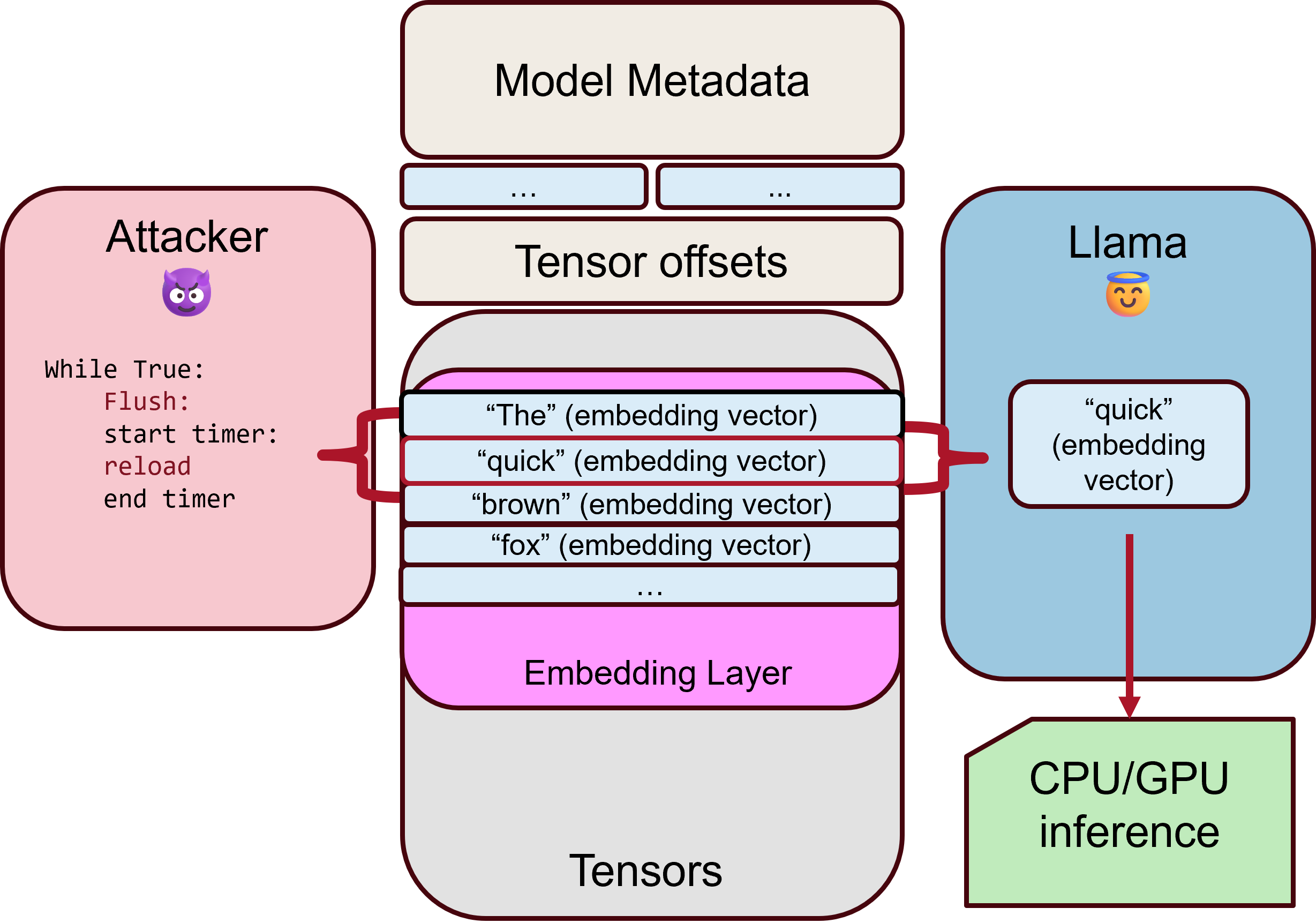}
    \caption{\label{fig:attack_overview}Overview of the \attack\ attack.}
\end{figure}

The \attack\ attack leverages cache side-channel techniques to extract tokens generated by large language models (LLMs) during inference. Figure~\ref{fig:attack_overview} illustrates the high-level attack flow. The adversary monitors memory access patterns in the embedding layer of the victim's LLM process by utilizing the Flush+Reload side-channel technique. This method detects when specific embedding vectors—corresponding to unique tokens—are accessed and cached. By observing these patterns, the attacker can reconstruct the token-by-token outputs of the victim's LLM inference.

In a typical scenario, the victim runs a process to access an LLM hosted on shared hardware, such as in a cloud environment. The LLM's inference operations are executed on an attached GPU, while both the attacker and victim reside on the same CPU. Importantly, we suspect that the attack capitalizes on unified CPU/GPU memory introduced in Nvidia Cuda 8, which ensures data coherency and allows embedding vectors to reside in CPU memory, making them susceptible to side-channel monitoring.

The attack proceeds in the following steps:

\begin{enumerate}[noitemsep,topsep=0pt,leftmargin=*]
    \item \textbf{Setup:} The attacker co-locates a malicious process on the same physical CPU as the victim's process, enabling access to the shared CPU memory space.
    \item \textbf{Calibration:} The attacker identifies the embedding layer's memory locations within the model file (e.g., GGUF file) using metadata and calculates the offsets for target tokens.
    \item \textbf{Flush:} The attacker uses the \texttt{clflush} instruction to evict specific memory addresses corresponding to embedding vectors from the cache.
    \item \textbf{Monitor:} During the victim's LLM inference, the attacker measures memory access times to detect cache hits. A cache hit indicates that the victim accessed the embedding vector for a specific token.
    \item \textbf{Inference Reconstruction:} The attacker correlates detected cache hits with token IDs to reconstruct the output of the victim's LLM inference token by token.
    \item \textbf{Iteration:} The attacker repeats this process to monitor multiple tokens, optimizing the trade-off between detection reliability and vocabulary coverage.
\end{enumerate}

%%%%%%%%%%%%%%%%%%%%%%%%%%%%%%%

%%%%%%%%%%%%%%%%%%%%%%%%%
\paragraph{Threat Model}
%%%%%%%%%%%%%%%%%%%%%%%%%
We assume the target system is free of any software/hardware vulnerability and logical protections such as sandboxing through process and VM isolation are in place. We assume co-location with the victim, where the adversary has \textit{software only} access to the processor or memory system. Another restriction we place is that the attacker does \textbf{not} have access to the victim GPU and only the CPU memory space needs to be physically shared while logically still isolated. Further, the attacker cannot run low-level physical side-channel or fault attacks, e.g. EM faults or monitor the victims system using any physical means, e.g. by attaching an oscilloscope or probing. Indeed, the attacker does not need physical proximity only co-location in the CPU memory space with the victim.  

While the use cases of LLM's are constantly evolving there are number of scenarios where the proposed attack can be mounted. As it stands LLMs are currently being integrated across our computing infrastructure, e.g. on websites, in mobile assistants, in corporate software. For instance, 
\begin{itemize}[noitemsep,topsep=0pt,leftmargin=*]
    \item a consumer facing application enabled by an LLM backend may become the victim to a co-located adversary; 
    \item in a corporate setting where shared servers run LLMs an employee might target others eavesdropping on their LLM queries
    \item by extending cache monitoring to a browser setting, a browser tab might spy on LLM queries on the local desktop.
\end{itemize}

\paragraph{CPU/GPU Memory Coherency in Nvidia Cuda 8+.}
The \attack\ attacker runs along with the victim on the CPU monitoring the victim's (CPU) memory accesses. This begs the question: since the actual inference is run on the GPU where do the actual token embedding tables physically reside? It is natural to expect the token embedding vectors to be loaded into the GPU memory, since the GPU is used to compute the token encodings and decodings at the beginning and end of LLM inference, respectively. If so, how can the attacker that is isolated to the CPU monitor accesses on the GPU memory? 

With the introduction of Nvidia Cuda 8 in 2017 \cite{cuda8}, many new features were introduced to simplify CPU/GPU coding and to improve the efficiency of memory management. Unified memory \cite{unified_memory} that can be freely accessed from both domains was for the first time extended to allow the GPU code to oversubscribe, i.e. allocate more than what is physically available on the GPU, similar to the way we use virtual memory in CPUs. In addition, the introduction of page faults on the GPU memory ensures \textit{CPU/GPU data coherency} even without explicit synchronization (and costly waits). This means as long as the embedding vectors are kept in unified memory, synchronized copies are kept on both devices due to data coherency. Hence in our threat model, the attacker can, by running flush+reload on the local CPU memory, monitor accesses to the embedding tables on the GPU memory thanks to coherency and cacheable tables.  

\paragraph{Token Representation and Embedding Layers.}
In natural language processing, words or tokens are typically represented as high-dimensional, sparse vectors. Embedding layers transform these tokens into dense vectors that capture semantic relationships~\cite{bengio2000neural}. In the embedding layer, each row corresponds to the embedding vector of a unique token.

During inference, when an LLM processes input text, it retrieves the embedding vectors for each token from the embedding layer~\cite{vaswani2017attention}. These vectors are then fed into subsequent layers of the model to generate contextually appropriate responses. The embedding vectors are frequently accessed during both the input processing and generation phases.

\paragraph{Memory Access Patterns in Embedding Layers.}
Accessing embedding vectors involves reading specific memory locations in the embedding layer corresponding to the input tokens. This process generates distinct memory access patterns that can be exploited by side-channel attacks. The embedding layer, a large and dense data structure, causes significant changes in the cache state, especially when multiple tokens are processed sequentially.

LLMs often have vocabularies with up to 128K tokens, resulting in embedding vectors that occupy significant memory space. The large size of these matrices frequently leads to rapid eviction of cache lines, making it challenging for attackers to monitor specific tokens without missing cache hits.

\paragraph{Challenges in Monitoring Embedding Layers.}
The size and complexity of embedding layers in large LLMs pose significant challenges for cache side-channel attacks. Specifically:
\begin{itemize}[noitemsep,topsep=0pt,leftmargin=*]
    \item \textbf{Cache Eviction:} The embedding matrices' massive size leads to rapid cache line turnover, increasing the likelihood of missing cache hits for monitored tokens.
    \item \textbf{Scalability Trade-offs:} Monitoring a broad range of tokens increases potential vocabulary leakage but raises the risk of cache misses due to eviction. Conversely, focusing on fewer tokens improves detection reliability but limits the scope of the attack.
    \item \textbf{High Dimensionality:} The dense representation of tokens in embedding vectors complicates distinguishing between tokens based on cache access patterns.
\end{itemize}

\subsection{Detecting Accesses to The Model}

To establish a foundation for detecting memory accesses in the embedding layer, we began by replicating the Flush+Reload side-channel methodology from \cite{yarom2014flush}. The goal of these initial experiments was to measure the latency difference between cache hits and cache misses when accessing specific addresses in a GGUF (GPT-Generated Unified Format) file. This calibration step was important for distinguishing between cache states in subsequent experiments involving token leakage.

We utilized the clflush instruction from Intel’s instruction set to explicitly flush specific memory addresses from the cache. By performing subsequent memory accesses to these flushed addresses, we measured access latency using the high-resolution RDTSC timing register. These measurements provided clear distinctions between cache hits (addresses retained in cache) and cache misses (addresses evicted from cache).

Our experiments yielded two distinct distributions of access times, corresponding to cache hits and cache misses. Cache hits consistently exhibited significantly lower latency compared to cache misses, allowing us to define a reliable threshold for differentiating between the two states.

To minimize false positives, a threshold value that captured 99\% of cache hit latencies, plus one standard deviation, could be a good selection. This choice ensured that the threshold fell solidly between the two distributions, effectively eliminating overlap and allowing accurate detection of cache states without misclassification.

% TODO: Add example of how tiny C code change results in 4-1 bit flip [DONE]

\begin{figure}
    \centering
    \includegraphics[width=\columnwidth]{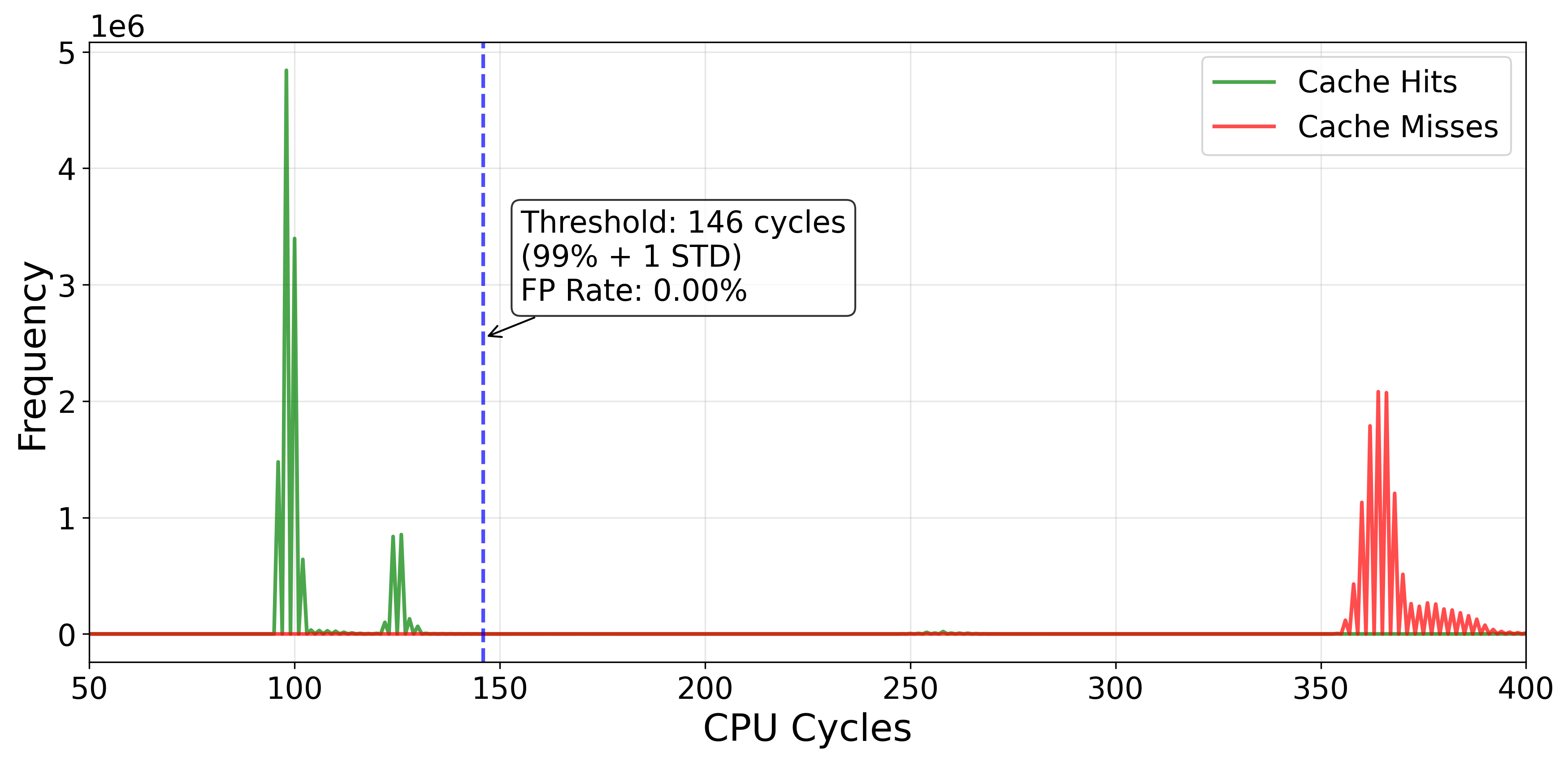}
    \caption{Calibration experiment demonstrating timing differences between cache hits (100 cycles) vs. cache misses (370 cycles)}
    \label{fig:calibration}
\end{figure}

The graph in Figure \ref{fig:calibration} illustrates the latency distributions for cache hits and misses, along with the selected threshold. This threshold optimization allowed for robust detection of cache hits in our calibration experiments. 

\subsection{Noise and Core Affinity in Detecting Memory Accesses}
\label{sec:Noise}

In this stage of the study, we examined the noise introduced when monitoring cache hits for memory accesses to a specific location in the GGUF file. To do this, we ran two separate processes: one actively reading from a particular memory address within the GGUF file, and the other monitoring cache hits around that address using Flush+Reload. Note that we have not introduced inferencing, or actually running the model - we are simpy reading bytes from the model file from another process. The goal was to evaluate the impact of system noise, such as prefetcher activity, and to explore the differences in behavior when processes were assigned to cores with varying levels of cache sharing.

Two configurations were considered for the core affinity of the processes:

\begin{itemize}[noitemsep,topsep=0pt,leftmargin=*]
\item Sibling Cores: The processes were placed on sibling cores that share the L2 and L3 caches but have separate L1 caches.
\item Same Core: Both processes were run on the same core, sharing the L1, L2, and L3 caches.
\end{itemize}

By monitoring latency distributions in these configurations, we sought to identify patterns and sources of noise in the observed cache hits.

\begin{figure}[!h]
    \centering
    \includegraphics[width=\columnwidth]{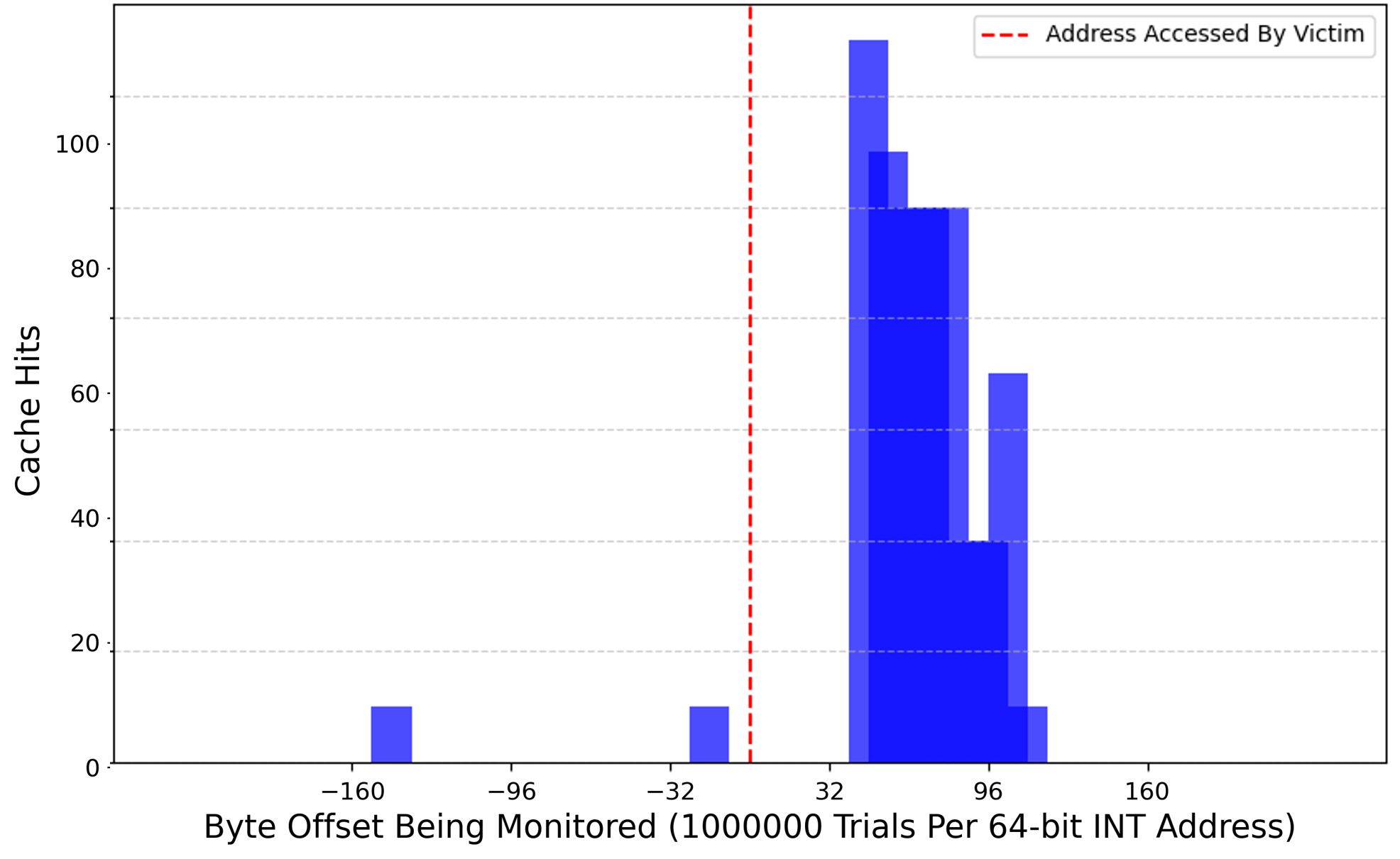}
    \caption{Detecting cache hits with Flush+Reload from addresses surrounding byte-address being accessed by a separate process on a sibling core}
    \label{fig:cache_monitor_sibling}
\end{figure}

\paragraph{Results on Sibling Cores.} When the processes ran on sibling cores, we observed cache hits primarily around (but not directly at) the address being accessed by the reading process, as seen in figure \ref{fig:cache_monitor_sibling}. This result suggests that while the monitored address was not explicitly reloaded into the cache by the sibling core, the shared L2 and L3 caches contributed to hits in nearby addresses, potentially influenced by speculative execution or prefetching mechanisms.

The observed cache hits followed a distribution centered near, but not exactly on, the target address. This behavior indicates a degree of noise around cache access detection when processes share only the higher-level caches.

\begin{figure}[!h]
    \centering
    \includegraphics[width=\columnwidth]{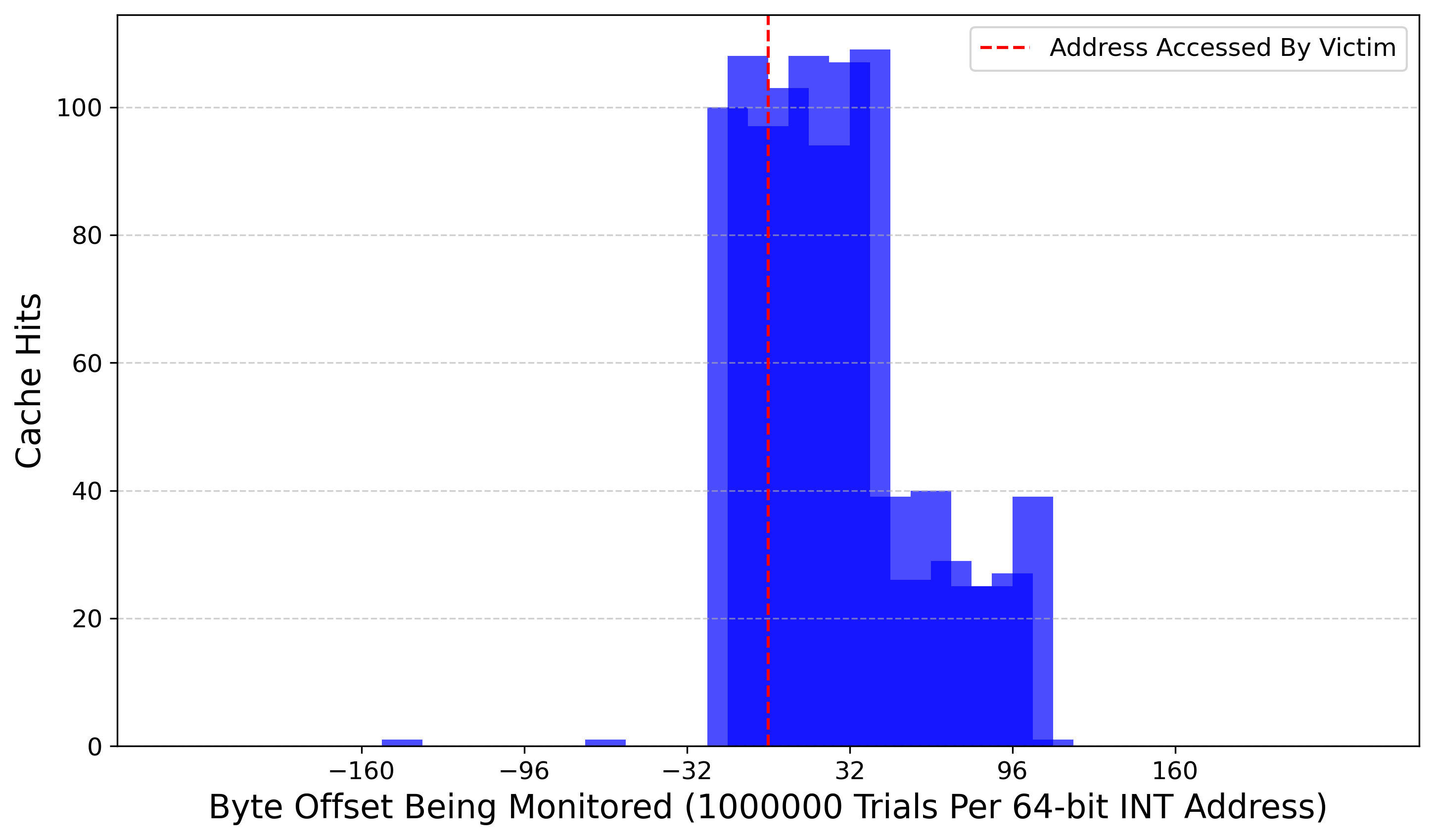}
    \caption{Detecting cache hits with Flush+Reload from addresses surrounding byte-address being accessed by a separate process on the same core}
    \label{fig:cache_monitor_same}
\end{figure}

\paragraph{Results on the Same Core.} When both processes were placed on the same core, the pattern shifted as seen in Figure \ref{fig:cache_monitor_same}. Cache hits occurred both at the exact memory address being accessed and in the surrounding addresses. Interestingly, this distribution filled in the gap observed in the sibling-core configuration, suggesting that the L1 cache's finer-grained access patterns enhanced the detection of memory accesses. The nearly perfect alignment of cache hits with the actual access confirmed that running both processes on the same core offers a more precise mechanism for detecting memory activity.

\subsection{Where to Monitor GGUF to Leak Tokens}
\label{Sec:GGUF_monitor_loc}
A GGUF file consists of four distinct sections:
Header, Metadata \& Tensor Info, Embedding Layer,
and Other Tensors

% \begin{itemize}[noitemsep,topsep=0pt,leftmargin=*]
% \item Header
% \item Metadata \& Tensor Info
% \item Embedding Layer
% \item Other Tensors
% \end{itemize}

The Metadata \& Tensor Info section of the GGUF file provides information required to locate and monitor specific tokens. This section, among other things, contains the byte offset of the embedding layer, the size of the embedding layer, and the total number of tokens in the model. 
To determine the byte offset of a specific embedding vector corresponding to a token, we use the following equation:
\begin{equation}
\text{Offset}_{\text{token}} = \text{Offset}_{\text{embedding\_layer}} + (\text{Token ID} \times \text{Size}_{\text{token}})
\label{eq:embedding_offset}
\end{equation}

\begin{itemize}[noitemsep,topsep=0pt,leftmargin=*]
    \item $\text{Offset}_{\text{embedding\_layer}}$: Byte offset of the embedding layer, obtained from the Metadata \& Tensor Info section.
    \item $\text{Size}_{\text{token}}$: Size of each token’s embedding vector, calculated as:
    \begin{equation}
    \text{Size}_{\text{token}} = \frac{\text{Size}_{\text{embedding\_layer}}}{\text{Number of Tokens}}
    \end{equation}
    \item $\text{Token ID}$: The unique identifier of the token.
\end{itemize}

\begin{figure}
    \centering
    \includegraphics[width=\columnwidth]{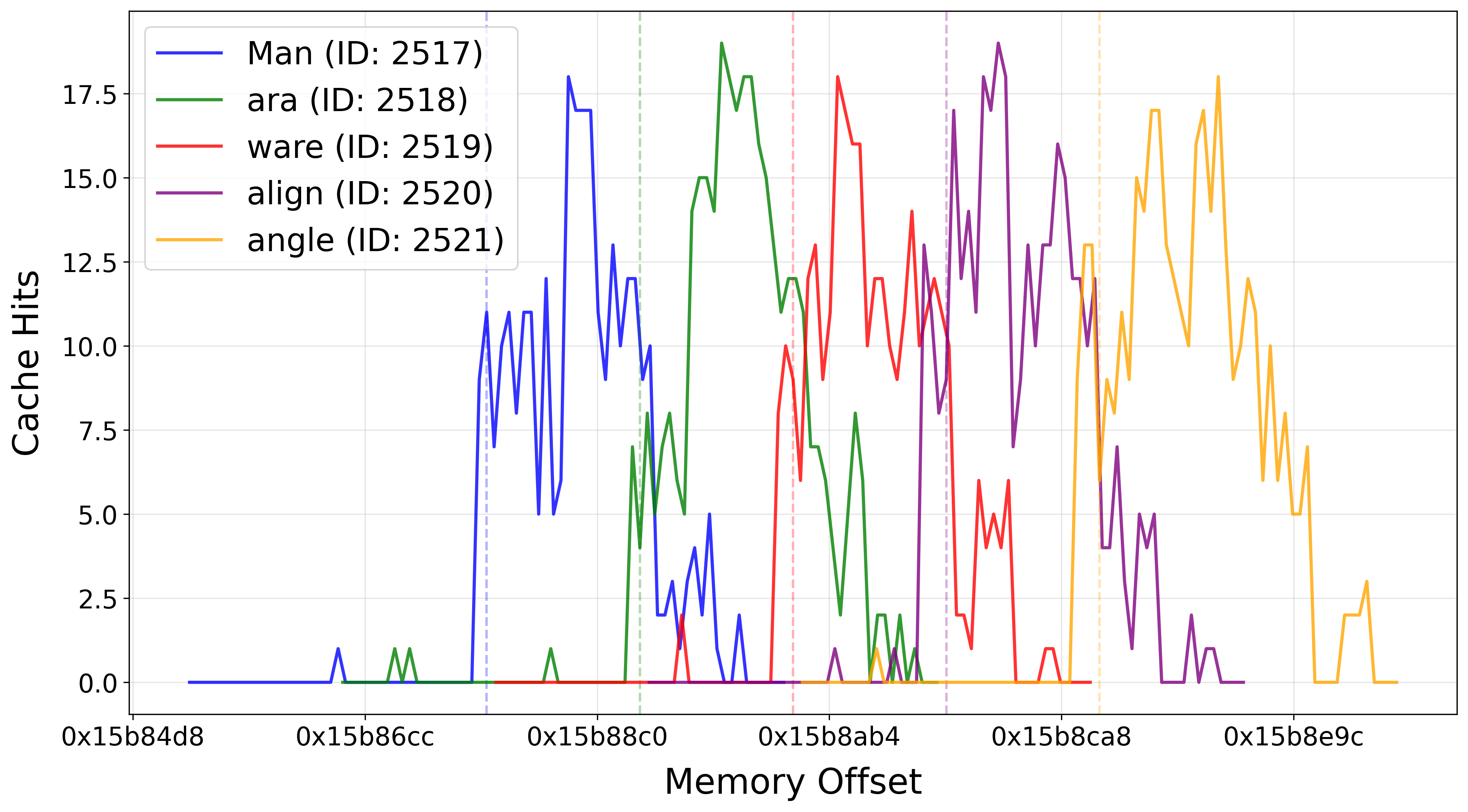}
    \caption{Detecting distributions of cache hits on embedding vectors for various tokens from a separate process on a sibling core}
    \label{fig:multi-token}
\end{figure}

This equation allows identification of the memory location corresponding to any given token in the model. By monitoring these computed offsets in the GGUF file, an attacker can observe cache hits for specific embedding vectors when tokens are processed by the model. 
To demonstrate this effect, we experimented with prompting the Llama model by Meta with only a single token, and monitored for cache hits where we expect the embedding vector to be, along with nearby locations. Just like in Section \ref{sec:Noise}, we expect a distribution of cache hits around the embedding vector when a particular token is used. We repeated this experiment for 5 tokens and graphed the results in Figure \ref{fig:multi-token}. You can see clear differences between cache hits to various tokens, indicating the first CPU cache side-channel to leak tokens from LLMs to our knowledge. 

\subsection{Round-robin Monitoring of Multiple Tokens}

While monitoring a single token offset provides a direct demonstration of the Flush+Reload technique, practical attacks often require tracking multiple tokens to infer richer linguistic information. Extending our approach to a broad set of tokens introduces both scalability and timing considerations, given the large size of LLM vocabularies and the limited time window before accessed cache lines are evicted.

To parallelize token monitoring, we first identify the token IDs of interest from the model’s vocabulary. Once we have these IDs, we convert each into its corresponding offset within the embedding layer. This conversion follows the same process described in Section \ref{Sec:GGUF_monitor_loc}. This yields a list of absolute offsets, one per token to be monitored.

With the set of offsets in hand, our attacker process—pinned to a sibling core of the CPU running the LLM—iterates over each target token’s memory location in a rapid, round-robin fashion. For each token offset, the attacker executes the following steps:

\begin{enumerate}[noitemsep,topsep=0pt,leftmargin=*] \item \textbf{Access}: Load the cache line at the computed offset in the GGUF file. 
\item \textbf{Time Measurement}: Record the time required to complete this load using a high-precision timing source such as the 
\texttt{RDTSC} instruction. 
\item \textbf{Flush}: Immediately flush the same cache line using 
\texttt{clflush}, ensuring the line is evicted from all levels of the cache hierarchy. 
\item \textbf{Record}: If the token access is within a specified time constraint, record the cache hit with as minimal overhead as possible.
\item \textbf{Next Token}: Continue round-robin to the next token to Flush+Reload.
\end{enumerate}

If the recorded access time is below a predetermined threshold (e.g., 200 CPU cycles), the attacker concludes that the token was recently accessed by the model and hence ``hit'' in the cache. This hit is then logged, mapping a low-latency read to the corresponding token ID. To minimize overhead during detection events, we refrain from conducting extensive processing or disk I/O until after the monitoring loop completes. This strategy reduces the risk of missing subsequent cache hits on other tokens while still enabling robust token reconstruction.

\paragraph{Overhead During Token Hit} We define the overhead for a token hit as the compute time between when a token is detected and when the round-robin Flush+Reload can continue. For example, if a token is detected, that token can be written to a file before we continue to monitor for more tokens - but because writing to a file can take a decent amount of time in the context of Flush+Reload, subsequent tokens may be missed while the attack is writing to the file. 

Alternatively, an attacker may want to change the tokens she is monitoring based on the token received.  In this case, there is a non-negligible amount of compute time taken between tokens to understand the context and select a new set of tokens to monitor, and tokens may be missed during this selection time. Thus, we wanted to quantify how many tokens you can expect to miss versus the overhead compute time between tokens. 

\begin{figure}[h!]
    \centering
    \includegraphics[width=\columnwidth]{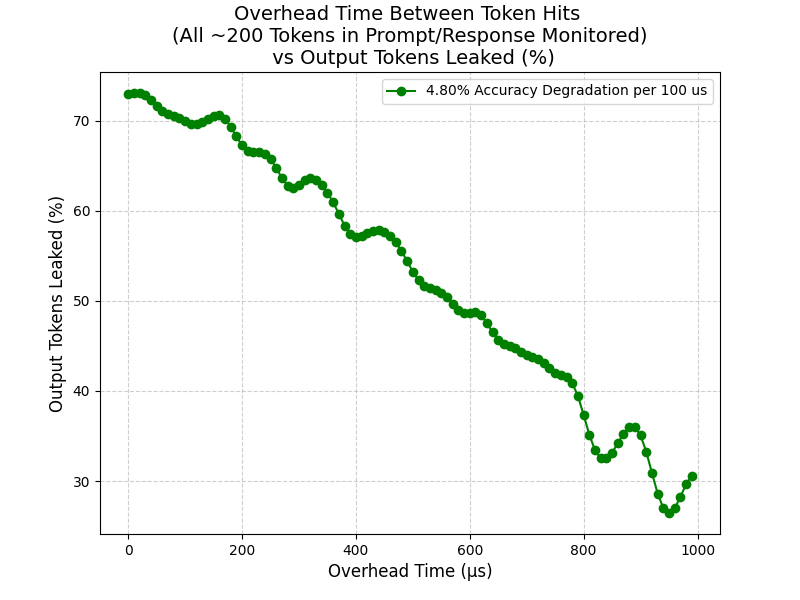}
    \caption{Tracking the output leakage of an LLM vs the overhead time between token hits}
    \label{fig:Overhead}
\end{figure}

\begin{figure}[ht]
    \centering
    \begin{tcolorbox}[
        enhanced,
        width=\columnwidth,
        boxrule=0.5pt,
        colback=white,
        arc=3mm,
        title=Chat Interaction Example Involving Sensitive Data,
        fonttitle=\bf series]
    \begin{minipage}{\columnwidth}
        % User message
        \begin{tcolorbox}[
            enhanced,
            boxrule=0pt,
            leftrule=3pt,
            colback=gray!10,
            colframe=gray!40,
            arc=1mm,
            width=\columnwidth]
        \textbf{User:} Please write a script that checks the value of my Bitcoin wallet. 
        
        My API key is: ``b81132e3-2129-42e3-91f4-d8b64c93fbec''. Please only give me the code
        \end{tcolorbox}
        
        \vspace{2mm}
        
        % Assistant response
        \begin{tcolorbox}[
            enhanced,
            boxrule=0pt,
            leftrule=3pt,
            colback=blue!10,
            colframe=blue!40,
            arc=1mm,
            width=\columnwidth]
        \textbf{Llama:}
        \begin{verbatim}
import requests
def get_bitcoin_balance():
    api_key = 
    "b81132e3-2129-42e3
    -91f4-d8b64c93fbec"
    
    response = requests.get(
    f"https://pro-api.coin
    gecko.com/api/v3/coins/bitcoin?
...
        \end{verbatim}
        \end{tcolorbox}
    \end{minipage}
    \end{tcolorbox}
    \caption{Example of a chat interaction showing potentially sensitive information being shared in a prompt.}
    \label{fig:chat-example}
\end{figure}

We ran a small 100M parameter model using a CUDA backend on an Nvidia 2070S GPU, and monitored ~200 tokens in a round-robin for a given prompt. Its important to note that we selected the ~200 tokens for this initial test based on what we would see in the output of the prompt, so assuming that no tokens are missed in the round-robin, they would all be captured by the attacker. We tracked how many of the ~200 tokens were received for each trial and recorded the accuracy of the response vs the overhead. The results as seen in Figure \ref{fig:Overhead} indicate there is roughly a 5\% loss token leakage for every 100~{\textmu}s of overhead during a token hit. This may vary between models or platforms, but ultimately it drove our decision to keep the overhead as minimal as possible to maximize the token leakage. 

\paragraph{Max Number of Tokens in Round Robin} Finding the right balance between coverage (the breadth of tokens monitored) and responsiveness (the likelihood of catching a given token when it appears) is crucial. On one extreme, monitoring a very small number of tokens—say, only five—enables the attacker to cycle through them rapidly, reducing the chance of missing a cache hit. If any of those five tokens is produced by the model, it is more likely to be detected, as the round-robin loop can return to each token’s offset before the embedding vector is evicted from the cache. However, such a narrow focus dramatically limits the range of information that can be leaked. With so few tokens under observation, even a near-perfect detection rate yields minimal insight into the victim’s secret prompt or text.

On the other hand, expanding the monitored set to a thousand tokens or more improves coverage, offering the potential to leak a broader range of linguistic content. However, this greater coverage comes at a cost. With a large set of tokens, the round-robin scanning takes significantly longer, increasing the time to return to any given token offset. If the model accesses a token at offset 10 in a list of 1000 tokens, but the attacker is currently monitoring offset 22, the loop must cycle through nearly a thousand tokens before returning to offset 10. By that time, the original token’s data is likely to have been evicted from the cache, resulting in a missed detection. The larger the monitoring set, the higher the probability that a given token access will go undetected.

This trade-off suggests the existence of an optimal set size—a number of tokens that is large enough to capture meaningful linguistic detail but not so large as to exceed the cache retention window. To identify this sweet spot, we conducted experiments using known prompt outputs and varying the number of monitored tokens. By trialing different set sizes and measuring the fraction of tokens successfully leaked, we identified an optimal point around 200 tokens. At this size, the attacker can cycle through the monitored tokens quickly enough to detect a substantial fraction of the accessed tokens while still covering a sufficiently broad portion of the vocabulary to yield informative leaks.

In practice, the optimal number of tokens to monitor will depend on factors such as the specific model’s embedding size, the system’s cache characteristics, and the temporal patterns of token generation. However, our findings highlight that neither extreme—focusing on a handful of tokens nor attempting to track thousands—is ideal. Instead, a moderate set size strikes the right balance, maximizing leakage by combining both coverage and timely detection.

\section{Leaking API Keys}
\label{sec:api_key_leakage}

In modern development workflows, it is not uncommon for end-users to embed API keys or other credentials directly into their LLM prompts, trusting the model to produce corresponding code snippets or instructions for integration into their applications. Unfortunately, this practice presents a critical vulnerability. Even under strict software isolation, a side-channel adversary leveraging \attack{} can recover such sensitive tokens as they appear in shared hardware memory during generation. Our demonstration, as outlined in Figure~\ref{fig:chat-example}, shows that \attack can reveal the full API key embedded in the LLM-generated code snippet, making it trivial for an attacker to assume the victim’s identity in downstream services.

\paragraph{Randomization and Entropy in UUIDs.} Universally Unique Identifiers (UUIDs), especially those adhering to versions defined in RFC 4122~\cite{leach2005rfc}, are intended to be statistically unique across space and time. A typical UUID, represented as a 36-character string (including 4 hyphens) such as b81132e3-2129-42e3-91f4-d8b64c93fbec, consists of a combination of randomly and fixed bits (depending on the UUID version) that provide a low-probability of collision and high entropy when interpreted as raw binary data. While the probability of two independently generated UUIDs colliding is astronomically small, from a side-channel perspective, this uniqueness and randomness significantly alter the distribution of tokens representing the UUID.

LLM tokenizers are trained predominantly on natural languages, resulting in token distributions that reflect common words and linguistic patterns. High-entropy strings like UUIDs do not conform to standard natural language distributions and are thus split into less frequently seen token sequences. For example, a string of hex characters b81132e3 may be tokenized into several low-frequency tokens rather than a few common ones. This property actually enhances side-channel leakage: the attacker only needs to round-robin monitor a sparse set of candidate tokens, as opposed to a large vocabulary of common substrings or words. The UUID’s unusual tokenization effectively reduces the search space once the attacker identifies a handful of rare tokens corresponding to the UUID’s uncommon substrings.

\paragraph{Contrast with Natural Language Tokens.} In natural language responses, tokens often represent partial words, punctuation, or common sequences encountered frequently in the model’s training data. Monitoring a broad swath of the vocabulary is both time-consuming and more prone to cache evictions. However, when the target is a random, high-entropy key the attacker can optimize their monitoring strategy by focusing on tokens that are unlikely to appear in everyday text. These tokens stand out in the embedding layer due to their infrequent usage and increased likelihood of hitting the cache at specific offsets when the model accesses them.
This probabilistic bias simplifies the attacker’s challenge: rather than monitoring a broad distribution of likely tokens (words like \texttt{def}, \texttt{and}, or \texttt{import} in Python code), the attacker can focus on a narrow set of memory locations that map to less common tokens. Hence, identifying the presence of UUID’s tokens is more straightforward, and extracting them as the model generates the code is significantly more efficient and reliable.

\paragraph{Implications for Credential Leakage.} The capability to leak random, high-entropy credentials such as UUID-based API keys highlights a severe privacy and security risk. The random distribution of characters in these keys, once thought to offer natural protection against brute-force attacks, becomes a liability for LLMs in the face of a direct hardware-based leak. An attacker who recovers an API key can impersonate the user or service, potentially gaining unauthorized access to cloud resources, databases, financial transactions, or other sensitive operations governed by the exposed API endpoints.

This threat is not limited to UUIDs. Any similar high-entropy secret—such as passwords, SSH keys, or cryptographic nonces embedded into LLM prompts—may be equally vulnerable. The convenience of using LLMs for code generation and instruction authoring must be balanced against the risk that these sensitive pieces of information can be inadvertently exposed through hardware-based side-channels.

\subsection{Token Monitoring}

\begin{figure}[h!]
    \centering
    \includegraphics[width=\columnwidth]{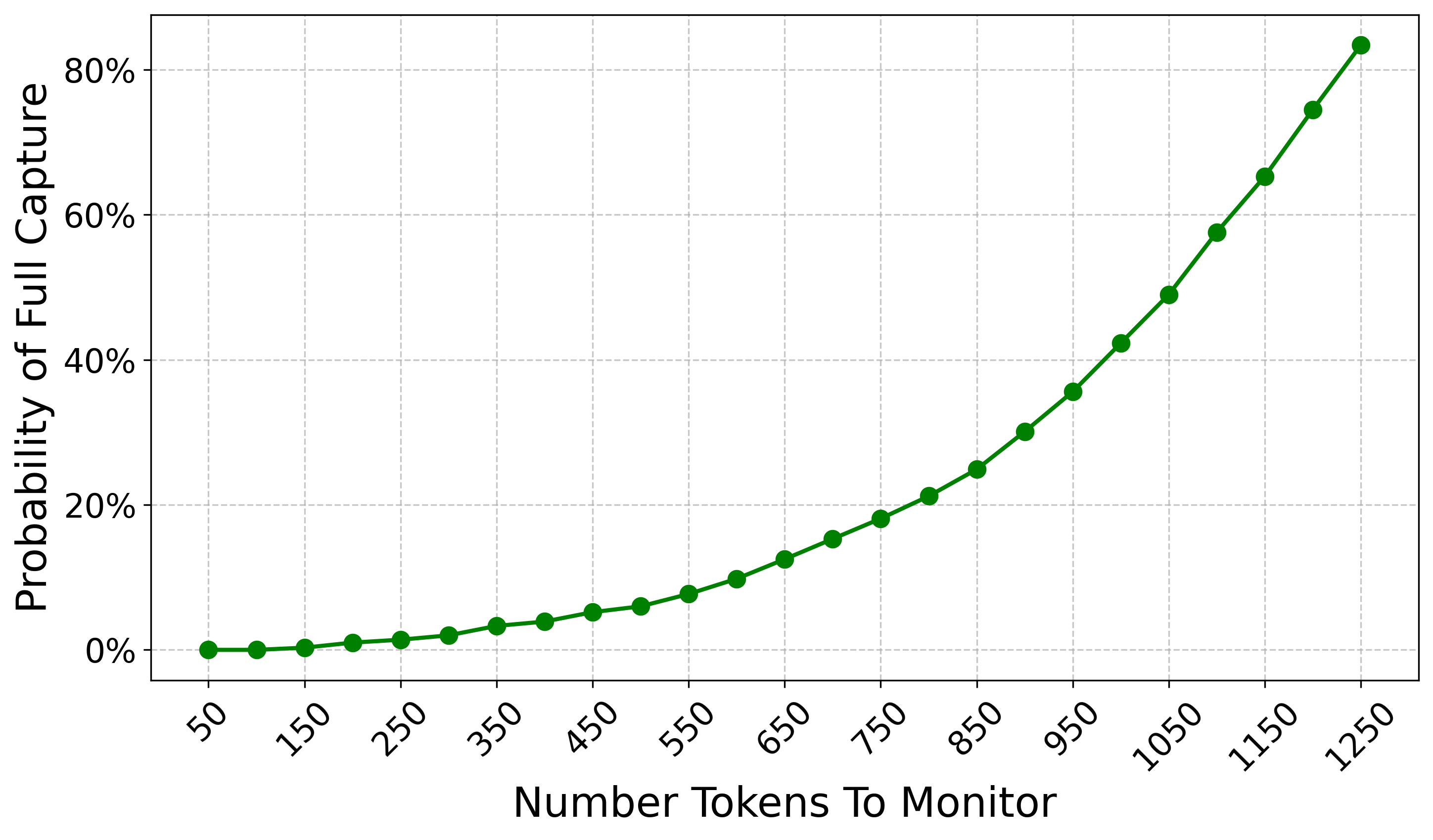}
    \caption{The probability of capturing an entire API key given a set number of tokens to monitor using \attack on a 128K token model}
    \label{fig:combos}
\end{figure}

Due to the high entropy of API keys, we need to monitor fewer overall tokens to capture the whole key compared to capturing an entire selection of English text. To determine an approximation for the number of tokens that we would need to monitor, we generated 100k random UUIDs and tokenized them based on the GGUF model. 
We then generated a new set of 100k UUIDs and determined how many of the tokens (sorted by frequency) we need to monitor for a high probability of leaking the whole key. Based on these tests, a clear relationship emerges between the number of monitored tokens and the probability of capturing all tokens in a high-entropy credential such as a UUID-based API key in Figure~\ref{fig:combos}.

Although the probability of full capture of 83. 4\% at 1250 tokens is significant, an attentive attacker can refine their strategy further. One potential tactic involves exploiting the tendency for high-entropy keys or credentials to recur multiple times over the course of an extended interaction. For example, a user might paste their API key into a prompt, request the LLM to integrate the key into a code snippet, and later ask the LLM to refactor the code which will result in a second leakage of the API key. This repetition naturally increases the total number of opportunities an attacker has to observe and fully reconstruct the key.

A practical attack strategy could proceed as follows: initially, the attacker monitors a broad selection of tokens (e.g., the top 250, providing a 86.3\% average coverage and a 1.4\% full capture probability). Even if the first occurrence of the API key does not yield a complete reconstruction, the attacker gains valuable intelligence. The missed tokens are now known to be among the less frequently used tokenized substrings of the credential. In subsequent occurrences of the key, the attacker can prune out subsets of tokens that have already been captured and rotate in tokens from a slightly different region of the embedding space—those less common tokens that were previously unmonitored.
This adaptive approach leverages the iterative nature of many user-model exchanges. Each repetition of the key offers a new opportunity to refine the monitoring set. Initially, the attacker might not know which tokens within the API key are hardest to detect. However, once partial information is gleaned, the attacker can strategically sample from a different selection of tokens in the next attempt, gradually filling in the missing pieces until the entire key is recovered. Over multiple appearances of the credential in the conversation, this rotation strategy converges on a complete set of tokens with high probability.
Notably, this method does not require the attacker to know anything about the content beforehand. The attacker simply monitors a broad set of tokens from the embedding space during the first exposure of the key. Identified tokens are “crossed off” from the search space. On the next exposure, the attacker replaces monitored tokens with those not previously observed. Repeating this process across two or three exposures can quickly push the full capture probability close to 100\%. In order to do this batch approach, we need to determine the maximum number of token that we can monitor in a round-robin accuracy loss.

\subsection{API Key Leakage Results}

To evaluate the practical effectiveness of our proposed attack strategy, we conducted experiments using a \texttt{llama.cpp}-based setup running on an Intel Comet Lake server CPU paired with an NVIDIA 2070S GPU. Although our experiments focused on \texttt{llama.cpp}, these findings generalize to other popular front-ends like \texttt{ollama}, since both utilize a similar GGUF model backend. We tested various configurations, monitoring different numbers of tokens and measuring both the theoretically predicted and actual number of user-model interactions required to fully reconstruct a UUID-based API key.

We began each experiment by selecting a set number of tokens for monitoring, informed by the coverage trade-offs discussed in the previous sections. The theoretical number of interactions represents an estimate derived from the statistical analysis of token coverage. The actual number reflects the empirical count of user queries and responses we needed before extracting the entire key in our end-to-end attack scenario. Factors such as subtle prompt variations, tokenization differences, and low-level microarchitectural timing all influence how closely the actual results match theoretical expectations.

Table~\ref{tab:api_key_leakage_results} shows the results from five experiments, each using a different number of monitored tokens and corresponding model configurations. As the number of monitored tokens increases, so does the theoretical probability of a full capture in fewer interactions. Our experiments confirm that by combining a well-chosen token set size with adaptive rotation strategies and multiple prompt exposures, an attacker can reliably leak the entire API key in a small number of interactions.

\begin{table}[h!]
\centering
\small
\begin{tabular}{l;{2pt/2pt}ccc} \toprule
\multicolumn{1}{l;{2pt/2pt}}{} & \multicolumn{3}{c}{\textbf{Interactions}} \\
\textbf{Set Size} & \textbf{to leak 50\%} & \textbf{to leak 80\%} & \textbf{to leak 100\%}\\
\midrule
50      & 1  & 7 & 66 \\
100      & 1  & 4 & 33 \\
200      & 1  & 1 & 18 \\
250      & 1  & 1 & 15 \\
300      & 1  & 1 & 12 \\
350      & 1  & 2 & 12 \\
400      & 1  & 2 & 9 \\
\bottomrule
\end{tabular}
\caption{Results from end-to-end attack on Llama models showing the number of interactions to leak different percentages of the full API key given the number tokens monitored by the attacker.}
\label{tab:api_key_leakage_results}
\end{table}

The results highlight that while perfect one-shot recovery may be improbable, and leak around 80\%-90\% of the API key on average, combining monitoring coverage with repeated prompt exposures and dynamic token selection yields a highly effective leakage strategy. Even in high-entropy domains like UUID-based API keys, modern hardware and inference pipelines offer numerous opportunities to piece together sensitive information, ultimately confirming the potency of cache side-channel attacks against LLM deployments.

%%%%%%%%%%%%%%%%%%%%%%%%%%%%%%%%%%%%%%%%%%%%%%%%%%%%%%%%%%%%%%%%%%%%%%%%%%
\section{Leaking Plain English}
\label{sec:english_leakage}
%%%%%%%%%%%%%%%%%%%%%%%%%%%%%%%%%%%%%%%%%%%%%%%%%%%%%%%%%%%%%%%%%%%%%%%%%%

While our initial experiments focused on leaking high-entropy tokens such as API keys, we next examined a setting where only plain English text was exchanged with the LLM. The underlying premise is that, despite lacking prior knowledge of the user's query or response content, an attacker can still infer a significant fraction of the generated English tokens by selectively monitoring the most frequently used words.

\paragraph{Building the Monitor Set.}
To construct a representative set of tokens, we utilized the Cornell Movie Dialogs corpus and extracted raw script lines for frequency analysis. By sorting words according to how often they appeared across the dataset, we identified the 1{,}000 most frequent tokens, which primarily consisted of articles, conjunctions, prepositions, and other common English words (e.g., \texttt{and}, \texttt{the}, \texttt{it}, \texttt{to}). Importantly, LLM tokenization often splits words into subword units, so we aligned each high-level word from our analysis with its corresponding GGUF token IDs to ensure the monitor set was accurate for our target model.
This reliance on high-frequency English tokens is directly tied to the well-known Zipf’s law, which describes the inverse power-law distribution of word frequencies in natural languages. Specifically, Zipf’s law posits that the frequency of the $n$-th most common word is roughly proportional to $1/n$. Interestingly, even random text generation can exhibit a similar rank-frequency relationship\cite{li1992random}, where the transformation from a word’s length to its rank effectively stretches an exponential distribution into a power law. This phenomenon implies that a handful of words often constitute a large proportion of written material. Thus, focusing on the 1{,}000 most frequent tokens captures a substantial fraction of likely English text, enhancing the feasibility of monitoring these tokens via Flush+Reload in a practical side-channel attack scenario.

\paragraph{Attack Setup with Quora Questions.}
We then chose several random questions from the Quora question set to simulate realistic user inquiries (\textit{e.g.}, \texttt{What is the best way to learn Python?}). These questions were passed as single-shot prompts to the \texttt{Meta-Llama-3.1-8B-Instruct-Q8\_0.gguf} model through \texttt{llama.cpp}. Concurrently, a separate attacker process pinned to a sibling core performed Flush+Reload on the embedding layer offsets corresponding to the 1{,}000 target tokens in a round-robin fashion. No additional context or prior knowledge about the query content was assumed.

\paragraph{Variable Token Set Sizes and Single-Shot Constraints.}
To assess the impact of the monitoring set size on leakage effectiveness, we repeated the experiment multiple times, varying the number of tokens from 50 up to 650. For each trial, the round-robin scan rate and timing thresholds were kept constant, allowing a fair comparison of how many tokens (in \%) were successfully detected under each condition. Unlike our API key analysis, these experiments were single-shot only: the prompt was provided once, the model response was generated, and no repeated or multi-turn dialogues were considered.  

\subsection{Results and Analysis}

\begin{figure}[ht]
    \centering
    \includegraphics[width=\columnwidth]{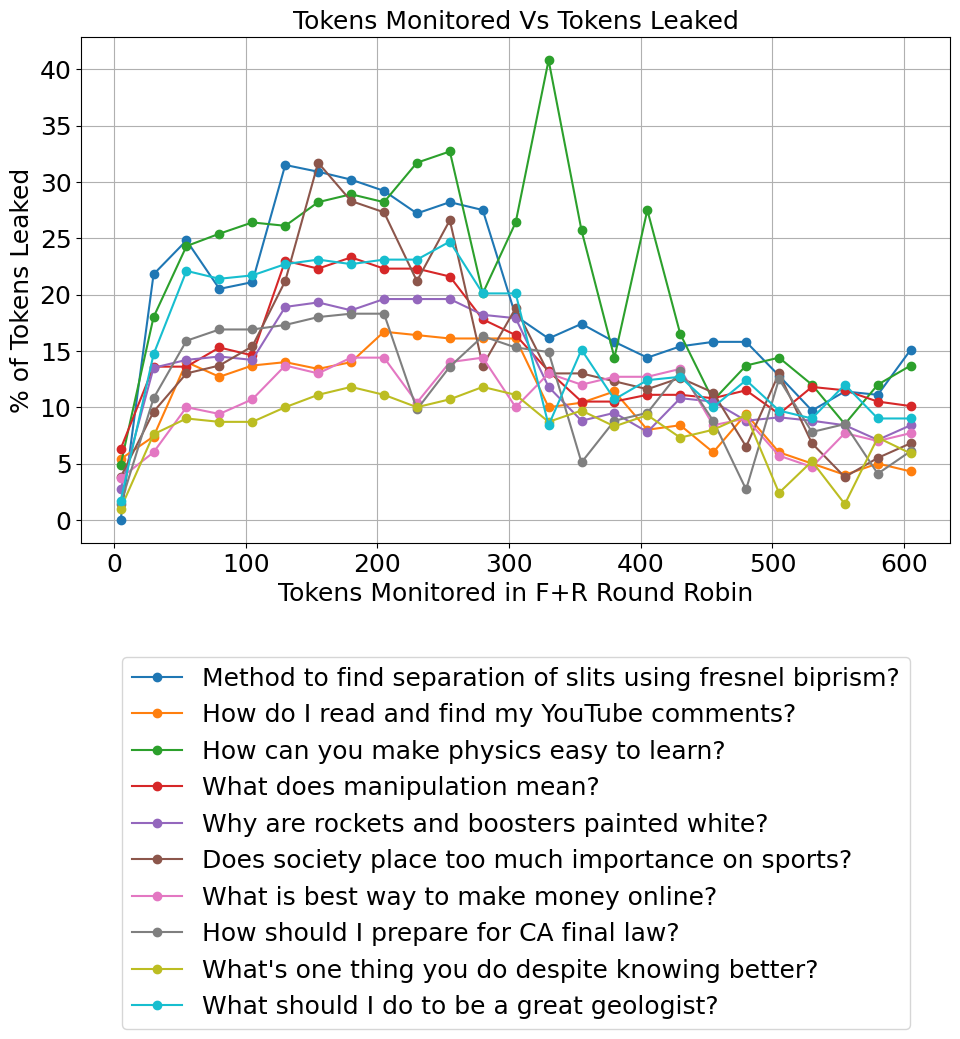}
    \caption{Percentage of plain English tokens successfully leaked vs.\ the number of monitored tokens (single-shot). A set size of 150--250 provides optimal coverage before round-robin misses dominate.}
    \label{fig:english_leakage_vs_monitored}
    \vspace{-0.1in}
\end{figure}

Figure~\ref{fig:english_leakage_vs_monitored} illustrates the observed percentage of tokens successfully leaked from the LLM responses as a function of how many tokens were monitored. Notably, increasing the monitor set size beyond roughly 150--250 tokens yielded diminishing returns. As the round-robin loop grew longer, the attacker process incurred greater delays when cycling back to any particular token offset. This increased latency led to higher eviction rates in the shared cache, causing the attacker to miss a proportion of token accesses.
Moreover, because the model’s response contained a mix of both frequent and infrequent tokens, focusing on the top English 150--250 tokens was sufficient to capture a substantial fraction of what the user was asking and the subsequent short answers. Monitoring a smaller set e.g. 50, missed more content while expanding beyond 250 produced excessive overhead and cache misses.  
\paragraph{Single-Shot vs.\ Multi-Shot Attacks.}
If the LLM reuses or repeats content across multiple turns of a conversation, the attacker can compensate for coverage limitations by rotating through different token subsets on each turn, as discussed in Section~\ref{sec:api_key_leakage}. Hence, although single-shot attacks already capture a significant portion of frequent tokens, repeated exposures would amplify the attacker’s ability to reconstruct a broader vocabulary. In practice, many user queries involve clarifications or follow-up questions, creating multiple opportunities to leak tokens across a longer conversation.
\paragraph{Implications for User Privacy.}
Our findings confirm that a naive assumption—that “only random data” is of interest to attackers—does not hold. Even plain English queries may reveal private or sensitive context \textit{e.g.}, personal details, medical information, or corporate secrets, when a large fraction of tokens are inferred. This exposure underscores the importance of adopting robust microarchitectural defenses or obfuscation strategies, even when only  English data is being exchanged.

\section{Further Improvements}
\label{sec:improvements}

The methodology outlined in this paper, while successful at leaking a considerable fraction of tokens through Flush+Reload, can be extended in several ways to further optimize leakage coverage and detection reliability. We briefly highlight three promising directions below.

\paragraph{Alternate Cache Side-Channels.}
Although Flush+Reload has proven to be effective and straightforward under shared-memory conditions, other microarchitectural attacks like Prime+Probe and Prime+Scope~\cite{kayaalp2016high, purnal2021prime} may enhance long-term visibility into victim accesses. In Prime+Probe, for instance, the attacker primes a cache set with its own data and later measures the time to probe that set, detecting evictions by the victim’s accesses without requiring shared pages. Prime+Scope narrows the granularity of probing to individual ways or slices, further improving signal quality. Both methods can potentially provide more stable observations and allow monitoring of a larger set of target tokens without risking frequent eviction misses. However, these strategies demand a deeper understanding of cache indexing and associativity parameters, adding complexity to the attack. Additionally, implementing Prime+Probe or Prime+Scope at scale can be intricate given hardware variations across CPU models.

\paragraph{Context-Aware Rotating Monitors.}
A second avenue for improvement is an adaptive scheme where the set of monitored tokens changes based on the partial information already gleaned from the victim’s text. As a simple example, detecting a sentence boundary token (e.g., a period) might trigger the attacker to pivot toward tokens likely to appear at the beginning of a sentence, such as capitalized words. More advanced mechanisms could use probabilistic language models or Markov chains to dynamically predict which tokens have the highest likelihood of occurrence, subsequently substituting seldom-used monitors with those that are contextually more probable. While such adaptivity requires additional computation per detected token, it may substantially boost the fraction of tokens recovered if carefully balanced against the added overhead.

\paragraph{Post-Processing with Language Models.}
Finally, even if the attacker misses some tokens or obtains partially corrupted sequences, modern LLMs themselves can be leveraged to perform automated text reconstruction. For instance, after collecting all leaked tokens and marking unknown or uncertain positions with placeholder symbols, an LLM or smaller specialized model could be asked to “fill in the blanks” with most probable missing tokens, using the known context. A subsequent refinement step could further rank alternative candidates for each blank token based on relative likelihood, improving the overall reconstruction. Although this post-processing adds computational expense, it is performed offline and therefore does not interfere with real-time side-channel measurements. Such a strategy can significantly enhance leakage probability, especially when the original text follows a syntactically rich structure or contains recognizable semantic cues.

\section{Countermeasures}

Mitigating cache side-channel attacks on LLMs requires a combination of hardware, system, and application level strategies. These strategies aim to reduce shared-resource contention, mask memory access patterns, and limit the attacker’s ability to correlate cache hits with token generation events.

\paragraph{Temporal and Spatial Randomization.}
Introducing noise and unpredictability into the LLM’s memory access patterns can substantially raise the bar for attackers. For instance, randomly accessing unused segments of the embedding layer during inference breaks the correlation between observed cache hits and actual token usage. By periodically injecting random read operations for tokens that are not currently generated by the model, the LLM can effectively drown out the signal the attacker relies on. Another approach would be to implement a “constant-time” strategy for LLM vector accesses, similar to those used in cryptographic libraries, where uniform access patterns thwart timing analysis. More research is needed to determine if this would be a practical approach, as additional accesses would cause latency to the inferencing.  Insuring that the attacker and victim processes do not share memory pages containing model parameters is another robust defense, however, it could be detrimental to performance. Enforcing process isolation at the hypervisor or container level, and avoiding shared libraries and models between untrusted tenants, can prevent cross-process Flush+Reload attacks \cite{zhang2012cross}. 

\paragraph{Hardware-Based Isolation and Partitioning.}
One effective mitigation involves using cache partitioning techniques, such as Intel’s Cache Allocation Technology (CAT) \cite{intelcat}, or similar hardware isolation features that limit the attacker’s visibility into victim cache lines. By isolating cache ways on a per-process or per-VM basis, these mechanisms ensure that eviction and timing differences observed by the attacker do not reveal meaningful information. Additionally, emerging hardware with strict spatial and temporal partitioning for shared resources can help prevent cross-core or cross-VM interference \cite{wang2019cacheguard}. Cache side-channel attacks like Flush+Reload often rely on page deduplication to create shared memory pages between attacker and victim \cite{gruss2015practical}. Disabling kernel same-page merging or deduplication features eliminates such opportunities. Although this may increase memory usage, the improved security posture would be significant in multi-tenant environments such as public clouds. However, this is generally not recommended due to the performance losses as a result.

\preto{\section}{\vspace{-2ex}}  % Reduces space before sections
\section{Conclusion}

This work introduces \attack, a practical demonstration that even state-of-the-art Large Language Models remain vulnerable to hardware-level side-channel attacks. Using cache timing variations in the embedding layer, we show that attackers can recover tokens used by the LLM, including sensitive keys and high-entropy credentials. Our experiments highlight how careful monitoring and strategic selection of tokens allow an adversary to overcome cache eviction and timing constraints, ultimately enabling the extraction of private data, such as API keys, from live inference sessions. Through extensive trials, we validated the feasibility of this approach on LLMs and explored how adjusting the number of monitored tokens influences both the breadth of data leaked and the reliability of detection.
Beyond simply exposing the vulnerability, our study contributes an improved understanding of the fundamental tension between model complexity, timing granularity, and cache resource sharing. We bring attention to the need for stronger defenses against microarchitectural threats and underscore that commonly suggested mitigations—such as disabling page deduplication or performing random unused token accesses—are likely necessary to preserve user privacy in multi-tenant environments. \attack\ serves as a reminder that the intersection of modern hardware design and AI models introduces new and subtle risks. To safeguard confidential interactions and private intellectual property, the community must pursue comprehensive hardware-software co-design solutions, isolation techniques, and adaptive obfuscation strategies that can effectively counter the evolving landscape of microarchitectural side-channels.

\clearpage  % Force a page break
\section{Disclaimer}
Andrew Adiletta's affiliation with The MITRE Corporation is provided for identification purposes only, and is not intended to convey or imply MITRE's concurrence with, or support for, the positions, opinions, or viewpoints expressed by the author. All references are public domain.

\section{Ethics considerations}
We have taken steps to ensure that all experiments were carried out safely and with the consent of the users of the said system. We plan to responsibly disclose the vulnerability to both Intel and Nvidia.

\section{Open Science Policy}
We plan on releasing code, data, and figures where allowed by the author's employer.

\bibliographystyle{plain}  % or another style like 'ieeetr', 'alpha', etc.
\bibliography{references}  % where 'references' is your .bib file name without extension

%-------------------------------------------------------------------------------

%%%%%%%%%%%%%%%%%%%%%%%%%%%%%%%%%%%%%%%%%%%%%%%%%%%%%%%%%%%%%%%%%%%%%%%%%%%%%%%%
\end{document}